\documentclass{article}
\usepackage[english]{babel}
\usepackage[left=2cm,right=2cm,top=2cm,bottom=2cm,bindingoffset=0cm]{geometry}
\begin{document}
 \begin{center}
 \Large
 \textbf{Q-balls and the chiral vortical effect} 
\normalsize
 \par Zakhar Khaidukov.\par

 Institute of Theoretical and Experimental Physics 
B. Cheremushkinskaya, 25, Moscow\footnote{khaidukov@itep.ru}
$$$$
\textbf{Abstract}

\par In this paper we try to show the possibility of  existence of an axial current directing along the axis of rotation in the presence of a non-topological soliton (the Q-ball). For  that purpose we'll analyse the case  when the Q-ball will be carring an  axial charge.  
And this condition will permit us to write terms in the effective action, which will be responsible for the existence of an axial current  and even for the chiral magnetic effect.
\end{center}

\large\textbf{1 Intoduction}
\normalsize \\

The chiral vortical and magnetic effects have  attracted a lot of attention recently.  For example, they were investigated in  $~\cite{1,2,3}$. As an example we may consider a medium consisting of chiral fermions, where conserved chiral charge $Q_{5}$ exists, and one can introduces chemical potential  $\mu_{5}$. If anyone mentions  the chiral magnetic effect, he understands the phenomenon of emergence of electric current $j_{el}$ directed along external  magnetic field B applying  to a chiral medium with a non-vanishing $\mu_{5}$
  $$\vec{j}_{el}=\frac{q^2}{2\pi^{2}}\mu_{5}\vec{B}, \eqno(1)$$
  where $\vec{B}$-magnetic field, q is the charge of the constituent fermion. 
  \par And by the chiral vortical effect one understands the phenomenon of emergence of the chiral current in the presence of nontrivial vorticity $\omega_{\mu}=\frac{1}{2}\epsilon_{\mu\nu\lambda\rho}u^{\nu}\partial^{\lambda}u^{\rho}$, where $u^{\nu}$ is 4-velocity of the medium.
 
$$\vec{j}_{5}=\frac{1}{2\pi^{2}}\mu_{5}^{2}\vec{\omega}\eqno(2)$$ and the axial current  directed along the axis of rotation.

The most interesting thing to note is that  coefficients in front of the vorticity also as in front of the magnetic field are connected with the chiral anomaly. There is the standpoint that, if we deal with a strongly interacting system, the coefficient  in  (1) in front of the  magnetic field  is not influenced by the  renormalization  and can be calculated explicitly [2]. In Son and Surowka's work [1] it was proved the possibility to come to the conclusion about existence of  chiral magnetic and chiral vortical effects only by consideration of hydrodynamic approximation. Moreover, in the work [1] it was not supposed that to obtain these results we need some special knowledge about the properties of the medium, but only thermodynamic parameters, currents and the energy-momentum tensor. It seems interesting to find a model in which it would be possible to check the results described above. In this paper we show that one of such possibilities is given by Q-balls [4], and we try to find the configuration of the field, which describes the state with the minimum energy for the given charge and an analog of the chemical potential. In this case it is possible to construct an effective action for Q-balls in an  external magnetic field and for a rotating Q-ball. From this action we may calculate chiral and  electric currents, and we can see that the coefficients in front of the rotational quantum number  \footnote{We mean that it plays the role similiar to the vorticity, because, in case of quantum system, the rotaion is described by rotaitional quantum number and by direction of the axis of rotation}
\\

\large \textbf{2 Description of introduction of the chemical potential}
\normalsize \\

The chiral vortical effect is connected  with  the chiral chemical potential, so, we need to understand how it may be introduced into the theory. Chemical potentials are introduced to simulate the effects associated with the presence of the medium. They are conjugate to corresponding conserved charges.
$$\delta H=\sum_{i}\mu_{i}Q^{i}\eqno(3)$$
\par From the thermodynamic point of view chemical potential is the energy required to add a single particle carring quantum numbers of corresponding charge  into the medium without doing any work. With this in mind the full Hamiltonian is read as
$$H =H + H_{int} + \delta H\eqno(4)$$	
\par If we  have the energy - charge equation, we may ask a question: " What is the ground state of the system for a given charge?"' The answer to this question will lead us to the concept of the Q-ball. The most surprising fact is that we are able to obtain such a ground state, even in the case of a strongly interacting system.\\

 \large \textbf{3 Q-ball soliton and chiral currents}
 \normalsize \\
 
In this section we discuss the construction of Q-balls in the field theory. We need to add some remarks connected with our work. First of all, if we decide to introduce axial charge, we must require conservation of the axial current. That means that we must put quark's masses to zero because of equation $$\partial_{\mu}j^{\mu5}=2im\bar{\psi}\gamma^{5}\psi.\eqno(5)$$ The model described in the work $\cite{5}$ allows us to avoid the problem associated with the mass of the scalar particle, we are talking about Nonlinear sigma model with vanishing mass parameter.
\par	As  it was mentioned above, the conception of Q-balls arises as the answer to the question about the ground state of the system carring conserved charge Q. In the case under consideration  it  is axial   charge $Q_{5}$ connected with U(1) axial symmetry. Let's study  a case of 4 dimensions with  action $S_{field} = \int{d^{4}x(T-V)}$, where T indicates free part of the action, and V is connected with interactions. For  better understending we try to  illustrate our ideas for particular case of abelian group only with $U(1)_{a}$ with action $S = \int{d^{4}x\partial_{\mu}\phi^{*}\partial_{\mu}\phi-V(|\phi|)}$. We are looking for solutions of the field equations in the form
$$ \phi= \sigma(r)exp(iwt)\eqno(6)$$	
For simple U(1) theory, in  the case of non-abelian group we will have some complications related with presence of group generators. Even in the simplest model there are several ways to find the soliton with the given charge and with the same value of $\omega$, and which is implemented depending on the type of potential V. Herein we describe a method, which, from our viewpoint, allows to identify the most important features for construction of an effective theory.
\par Let's construct the current from the action for the $U(1)_a$, it has the form 
$$J_{\mu}=-i(\phi^{*}\partial_{\mu}\phi-\phi\partial_{\mu}\phi^{*})\eqno(7)$$
 The charge is zero component of this current $Q = \int{d^3xJ_{0}}$, then we form the following expression:
 $$\epsilon=E_{f}-\lambda Q\eqno(8)$$
 Let's substitute (6) in   it , then the expression  becomes
$$\epsilon=\int{d^{3}x(\omega^{2}\sigma^{2}+(\partial_{i}\sigma)^{2}+V(\sigma)-2\lambda\omega\sigma^{2})}\eqno(9)$$

\par And now we need to find its extremum. This expression will allow us to find the ground state of the system. It seems a bit surprising that it's possible even in the case of strongly interacting systems. Condition $\frac{\partial \epsilon}{\partial \omega}$ gives us $\lambda=\omega$. Also, we can find the explicit form of function $\sigma(r)$. In the simplest case it is constant till some definite radius.
\par Let's analyze the obtained solution. We have find the ground state that having  charge $Q_{5}$. One can see that $\omega$ plays the role  similiar to the  chemical potential. If the charge is axial, so it is possible to find an expresision for  the chiral chemical potential.\\

\large \textbf{4 Rotation}
\normalsize\\

In  this section  of our work  we try to realize an idea of creation of a solution, which describes the rotating Q-ball.
Such well-developed solution is a very laborious task $~\cite{vol1}$ . Seeing that qualitative possibilities of  existence of axial and electromagnetic currents are interesting for us, for illustrative purposes, we focus on the simpler case of Q-vortex.\\
		\par Let's  introduce polar coordinates in the plane x,y. To take into account the rotation of the field configuration,  phase
$$\Phi=\sigma(r)exp(i\alpha),\eqno(10)$$ is to be represented in  form  $$\alpha=\omega t +N\phi,\eqno(11)$$
where $\phi$  is the polar angle. As we see, such substitution does not change  the density of the  Q-ball charge, due to the fact, that   $\partial_{0}\phi$=0, it's quiet clear, that the phase of field  $\Phi$ must change to a multiple of two pi when traversing a closed loop around the point $r=0$. Thus, we see, that N  has only integer values. Let's  seek minimum of field configuration at the given N and charge Q.
 \par Let's write the equation of motion
$$0=\sigma^{''}+\frac{1}{r}\sigma^{'}-\frac{N^{2}}{r^{2}}\sigma-\frac{dU(\sigma)}{d\sigma}+\omega^{2}\sigma \eqno(12)$$  
 Energy density per unit length is expressed as
 $$E=\pi\int{\omega^{2}\sigma^{2}+(\sigma^{'})^{2}+\frac{N^2}{r^{2}}\sigma^{2}+U(\sigma) dr^2}.\eqno(13)$$
If N=0, it is just the classical expression for the energy of the Q-ball. From the condition of energy finiteness we find that  $\sigma$  must vanish on the vortex core (r=0). Let us consider the asymptotic behavior of this equation: \\
When  $r \simeq 0$  $$\sigma=Ar^{N}+O(r^{N+1})\eqno(14)$$
and when $r->\infty$ 
 $$\sigma=\frac{B}{\sqrt{r}}exp(-\sqrt{(U^{''}(0)-\omega^2)}r).\eqno(15)$$ 
In addition, it can be shown, that the mechanical moment is quantized in units of charge: $J=\int{T_{0\phi}}drd\phi=NQ$. This predicating remains  correct even when the Q-ball is considering  $~\cite{vol1}$.
  \\
 
 \large \textbf{5 Chiral effects  and Q-ball}
 \normalsize \\
 
\par 
At the first step we consider a non-rotating Q-ball, which has a high charge, so high, that  its radius  $R\sim Q^{\frac{1}{3}} $  is much greater than the Compton wavelength  $E_{sol}^{-1}\sim Q^{-1}$. Then, in this limit, the Q-ball may be considered simply as a classical field configuration.

	\par Now let's study simple  U(1) model for the Q-ball. We need to understand how anomalies arise in this task.
For this we consider a system on scales, where fermionic degrees of freedom exist. Then the full Lagrangian may be expressed as
$$L=\int{d^{4}x\bar{\psi}i\gamma^{\mu}\partial_{\mu}\psi}+\int{d^{4}xL_{int fg}}+\int{d^{4}xL_{int f q}}, \eqno(16) $$ 
where term $L_{int fg}$  is responsible for interacting of fermions with "`gluon"'$\footnote{under the term "`gluon"' we understand all the other fields that are responsible for the formation of bosonic degrees of freedom } $ fields, and term    $L_{int f q}$    is responsible for interacting of fermions and the Q-ball. Effective action must be invariant with respect to the axial symmetry which acts on the fields as follows    $$\phi\to\exp(i\delta)\phi, \psi\to exp(iq\gamma^{5})\psi,\eqno(17)$$ where $\delta$ -is  corresponding charge of scalar field.$\footnote{for simplisity we will set it equal to 1, }$. To satisfy these requirements  we select an interaction  term as
  
     $$L_{intfq}=\frac{1}{2}\bar{\psi}\partial_{\mu}Ln(\frac{\Phi}{\Phi^{*}})\gamma^{\mu}\gamma^{5}\psi \eqno(18)$$ the most simple case that is  valid by symmetries, and we also include the interaction with U(1)-gauge field $q\bar{\psi}A_{\mu}\gamma^{\mu}\psi$.
    To obtain the final expression for the current we need to integrate over the "`gluon"' fields. In article  ~\cite{6} they explaine  the way of doing that.  Then, using the Fujikawa-Vergeles method $\cite{9}$, we find currents  under fermionic field  transformations
     $$\psi \to exp(i\gamma^{5}\alpha+i\beta)\psi\eqno(19),$$
     where $\alpha,\beta$ are independent parameters of transformation. As a result we will have two currents. The first one is 
an ordinary electromagnetic current: 
$$\partial_{\mu}j^{\mu}=-\frac{q}{4\pi^{2}}\epsilon^{\mu\nu\lambda\rho}\partial_{\mu}A_{\nu}\partial_{\lambda}\partial_{\rho}ln\frac{\Phi}{\Phi^{*}}\eqno(20)$$
On  larger scales we have no fermionic degrees of freedom, but the anomaly is still there. Thus we see, there is a macroscopic current 
$$J^{\mu}=-\frac{q}{4\pi^{2}}\epsilon^{\mu\nu\lambda\rho}\partial_{\lambda}ln\frac{\Phi}{\Phi^{*}}\partial_{\lambda}A_{\rho},\eqno(21)$$
if we insert  the equation for the Q-ball in it, we can see, that the electric current exists there, is 
 $$\vec{J}=-\frac{q}{2\pi^{2}}\omega \vec{B},\eqno(22)$$ it also has 
been shown in ~\cite{5}. One can see that within this theory we may come to  the conclusion of the existence of the effect similar to the chiral vortical effect. The fact is, that in the axial current 
 term

$$J_{\mu}^{5}=\frac{1}{16\pi^{2}}\epsilon_{\mu\nu\lambda\rho}\partial^{\nu}ln\frac{\Phi}{\Phi^{*}}\partial^{\lambda}\partial^{\rho}ln\frac{\Phi}{\Phi^{*}}\eqno(23)$$ arises.
  As we see, we can give rotation to the Q-ball by replacing  $\omega t\to \omega t + n\phi$, where we introduce the cylindrical coordinates, and angle  $\phi$ is the angle in the xy plane. In the expression for the axial current there is the commutator of two derivatives, and for any smooth function this commutator is equal to zero. And still, in case under consideration  we may prove that:  
$$ \frac{1}{4\pi}\oint\partial_{i}{\ln\frac{\Phi}{\Phi^{*}}}dx^{i}=n \eqno(24)$$ Using the Stokes' theorem we see  $\epsilon^{ijk}\partial_{j}\partial_{k}{\ln\frac{\Phi}{\Phi^{*}}}=4\pi\delta(x,y)n ,$ taking into
 account this relation, we find that the expression for the axial current becomes
  
$$J^{5z}=\frac{1}{2\pi}\omega n\delta(x,y). \eqno(25)$$
We integrate this current in the plane  x,y and  find that the axial current that is  directed along the axis of rotation and equal to$$J^{5z}=\frac{1}{2\pi}\omega n,\eqno(26),$$ must be there.
 Thus, in the simplest case, we can prove existence of the chiral magnetic effect and the effect similar to the chiral vortical effect.  \\

\large\textbf{6 General case}
\normalsize \\
In the simplest case  we succeed in fixing the existence  of an effect similar to the chiral vortical effect in the case of existence of the chiral magnetic effect,  and  the role of the chiral  chemical potential $\mu_{5}$  $\omega$ plays . But such discription is not accurate enough. The reason is, that, if the interaction term has the form $\frac{1}{2}\bar{\psi}\partial_{\mu}Ln(\frac{\Phi}{\Phi^{*}})\gamma^{\mu}\gamma^{5}\psi$, and then, as one can easily find, the existence  of the Q-ball is reduced to the appearance of  the "`chiral chemical potential"' in the  current, and it is independent of  the configuration of the field $\sigma$. But  symmetries  allow to write a much more general case of interaction.
Now we need to modify the action for the purpose  to take it into account. The most general term  we can write
$$+g\bar{\psi}\gamma^{\mu}\gamma^{5}f(....)\partial_{\mu}\zeta\psi,\eqno(27)$$
 where $\zeta$-describes the phase of the Q-ball, and g is the coupling constant, ànd f-scalar function, that is invariant with respect to the axial transformations.
Next, we define $f_{\mu}=gf\partial_{\mu}\zeta$. where $\zeta$-denotes the phase of the Q-ball. In order to obtain the chiral magnetic effect we also need to consider the interaction of fermions with U(1)    gauge field, to do this we add to the Lagrangian term
$q\bar{\psi}A_{\mu}\gamma^{\mu}\psi$. Now, using the method of Fujikawa-Vergeles $\cite{9}$ we will find 4-divergence of the axial and vector currents  $$\partial_{\mu}j^{\mu}=\frac{1}{2\pi^{2}}q\epsilon^{\mu\nu\lambda\rho}\partial_{\mu}A_{\nu}\partial_{\lambda}f_{\rho}\eqno(28)$$ 
$$\partial_{\mu}j^{\mu 5}=-\frac{1}{4\pi^{2}}\epsilon^{\mu\nu\lambda\rho}\partial_{\mu}f_{\nu}\partial_{\lambda}f_{\rho}\eqno(29)$$
out of (28) we can see the existence of a macroscopic electrical current
$$J^{\mu}=-\frac{q}{2\pi^{2}}\epsilon^{\mu\nu\lambda\rho}f_{\nu}\partial_{\lambda}A_{\rho}\eqno(30)$$ 

Let's consider the case of the mentioned above  non-rotating Q-ball, when  $\phi=\omega t$, and  if we  neglect with effects, associated with the boundary of the Q-ball, the current becomes 
$$\vec{J}=-\frac{1}{2\pi^{2}}qgf\omega\vec{B} \eqno(31)$$ where $B^{a}=\frac{1}{2}\epsilon^{abc}F_{bc},a,b,c=1,2,3, $  
   and we have the results of the preceding paragraph, connectd with  the chiral magnetic effect $\cite{6}$,  where  the value  $gf\omega=\mu_{5}$. plays
the role of the "`chiral chemical potential"'.
Let us now consider how changes in the action affect   the existence of the axial current 
We substitute the explicit dependence of the phase
 $\zeta=\omega t+n\phi,$ $\phi$- the polar angle. Then, the  axial current along the axis of rotation (in this case - z) is given by
$$J^{z5}=\frac{1}{4\pi^{2}}f^{2}g^{2}n[\partial_{x}\partial_{y}-\partial_{x}\partial_{y}]\phi\eqno(32)$$ and thus we again meet the situation that the commutator in (31) vanishes everywhere except at the origin of coordinates, so, the presence of the effect is determined by the behavior of function  f. If we take as f, for example, the function   $g\sigma^{2}$   with  asymptotic behavior  (14),  it is easy to see the axial current will not exist.\\

  \large \textbf{6 Conclusion}\\
  \normalsize\\
In this work we have considered the effective theory describing the origin of the chiral magnetic effect and the existence of a chiral current directed along the axis of rotation in presence of the axial Q-ball. It is shown that the presence of  the chiral magnetic effect can be fixed only on the basis of consideration of  symmetries.  As for chiral current directed along  the Q-ball axis of rotation its existence depends on the details of interaction.

\end{document}